\documentstyle{article}
\newcommand{\be}{\begin{equation}}
\newcommand{\ee}{\end{equation}}
\newcommand{\mb}{\mbox}
\newcommand{\fr}{\frac}

\newcommand{\p}{\partial}

\begin{document}

\title{Variational dynamics in open spacetimes}
\author{Jelle P. Boersma \\ ~ \\
Department of Applied Mathematics, University of Cape Town \\
Rondebosch 7700, Cape Town, South Africa}
\maketitle
\[
\]

\begin{abstract}

We study the effect of non-vanishing surface terms at spatial infinity
on the dynamics of a scalar field in an open
Friedmann-Lema\^{\i}tre-Robertson-Walker (FLRW) spacetime.
Starting from the path-integral formulation of quantum field theory,
we argue that classical physics is described by field configurations
which extremize the action functional in the space of field
configurations for which the variation of the action is well defined.
Since these field configurations are not required to vanish
outside a bounded domain,
there can be 
a non-vanishing contribution of a surface term to the variation of the
action.
We then investigate whether this surface term has an effect 
on the dynamics of the action-extremizing field configurations.
This question appears to be surprisingly nontrivial in the
case of the 
open FLRW geometry
since surface terms tend to grow as fast as volume
terms in the infinite volume limit.
We find that surface terms can be important for the dynamics of
the field at a classical
and quantum level,
when there are supercurvature perturbations.

\end{abstract}

\section{Introduction}

The idea that surface terms can be important when the Lagrangian 
method is applied to cosmology has been studied earlier in the 
context of spatially homogeneous but anisotropic models 
\cite{mmac} - \cite{ryan}.
In this case, a surface term appears when the Lagrangian is 
varied with respect to a spatially homogeneous metric 
perturbation, and the assumption of spatial homogeneity 
prevents the vanishing of this term when it is evaluated
on an arbitrarily distant compact two-surface. 
In most other cases where the variational approach is applied
to cosmology, surface terms are made to vanish trivially by 
evaluating only variations with respect to variables which 
vanish outside a bounded domain. 
The justification for this approach seems to be that one recovers
the `correct' field equations, which are the standard 
Euler-Lagrange equations.
In a cosmological context, this way of reasoning can be questioned, 
both from a theoretical and an observational point
of view.
From observations, it is not {\it a priori} clear which are the correct
equations of motion describing the dynamics of fields at length 
scales larger than the observable universe, and in different 
cosmological models. 
From a theoretical point of view, the relation between extremal 
action fields and classical physics has a natural foundation 
in quantum field theory. 
However, field configurations which vanish outside a bounded 
domain do not play a central role in quantum field theory, and 
this assumption may be questioned in a cosmological context 
when the spacetime itself is not bounded.

  In this paper we will study this situation by means of an
idealized model, which consist of a Klein-Gordon field in both
a spatially flat and a spatially open FLRW universe. 
Our motivation for studying a scalar field stems from the aim to 
keep our equations simple, and the possible importance of these 
fields in the description of the early universe.
We will concentrate on the open FLRW geometry, since this 
spacetime has some specific properties which allow 
surface terms to become important. 

One of these properties is that eigenfunctions of the spatial
Laplacian occur in two types.
First, there are eigenfunctions with eigenvalues exceeding
$\fr{1}{6}$ times the
spatial curvature, and these eigenfunctions are complete in the
space of square integrable functions \cite{bander}.
Second, there are eigenfunctions of the spatial Laplacian with
eigenvalues between zero and $\fr{1}{6}$ times the spatial curvature.
This last type of eigenfunctions cannot be square integrated, and
they are responsible for long-range correlations in a spatially
open universe \cite{lyth}.
In spite of the fact that these perturbations cannot be square integrated,
they may naturally occur in an open universe which is created in an
exponentially expanding false vacuum \cite{cohn, kaiser}, or
they may be generated during preheating \cite{ratra}.

Another important property of the open Friedmann-Lema\^{\i}tre-Robertson-Walker (FLRW) geometry is that a spatial
volume and the surface of its boundary grow at the {\em same} rate when the
infinite volume limit is taken.
The combination of large boundary surfaces and the presence of
long-range correlations in open spacetimes appears to have an
effect on the growth of surface terms at spatial
infinity in these spacetimes.

Besides the theoretical reasons which make the open FLRW 
spacetime an interesting object to study, the open FLRW 
geometry has gained relevance as a model for the observed universe, 
with observations favoring a relatively small value of the
density parameter \cite{ratra&peebles}.
Furthermore, progress has been made in describing the creation 
of an open FLRW universe from an exponentially expanding false 
vacuum (see, {\it e.g.}, \cite{bucher, sasaki}),
and the theory of perturbations 
in open FLRW spacetimes has been worked out in greater detail
\cite{lyth} - \cite{sasaki2}.

This paper is structured as follows. 
In section \ref{eap} we discuss the physical relevance of
action-extremizing field configurations, and we show that
surface terms can contribute to the variation of the action for
square-integrable perturbations.
In section \ref{ead} we decompose the scalar field perturbations
in terms of eigenfunctions of the spatial Laplacian, and we
discuss the occurrence of supercurvature modes.
The dynamics of the extremal action configurations is considered in section
\ref{dyn}, and we recover the usual equation of motion
for each perturbation component, with an additional
source term, which can be expressed in terms of a surface
integral which is evaluated at spatial infinity.
We show that this source term can be neglected in the case
where there are only subcurvature excitations of the scalar field, 
but it appears to diverge in the case where there are 
supercurvature perturbations.
Due to this divergence,  
extremality of the action can only be defined in the restricted 
phase-space of field 
perturbations for which surface terms are finite.
Depending on how one restricts the phase-space of field
perturbations, a nontrivial source term contributes
to the equation of motion for the extremal action
configurations.
In section \ref{q}, we consider the quantum correlation function 
of the
scalar field. In the case where there are supercurvature
perturbations, it is shown that the action functional is sensitive   
to degrees of freedom of the scalar field which
have zero $L_2$-norm. It therefore appears that
the correlation
function is not well defined, unless one adopts nontrivial 
constraints on the phase-space of the scalar field,
or one needs to include the zero-norm degrees
of freedom in the integration over paths.

\section{The extremal action principle}\label{eap}

In this section we briefly review the variational approach to classical
field theory. We then use arguments from quantum field theory 
to motivate a modified form of the variational method 
in a cosmological context. Surprisingly, it appears that
non-local interactions at a classical level can emerge from 
the underlying quantum theory with a standard expression
for the Lagrangian.   
While our explicit calculations involve only the simple case
of a scalar field, our arguments are relevant in a more general
field theoretical context, including
general relativity. We will come back to this point at the
very end of this paper.  

One way of describing the dynamics of a classical field is by
formulating a field equation. A specific solution of the field 
equation is determined by the boundary or periodicity conditions
which apply to the system.
It is of interest to note that the dynamics of the fields which
can be observed in nature are described by field equations which
act locally, while mathematical consistency does not require this.
Hence, the dynamics of classical fields has a local aspect, in the 
sense that the field equations involve only the field variables  
and derivatives thereof at each point.
Further, a particular classical field configuration is subject to
global constraints, which act in the form of boundary or
periodicity conditions.  
The work on this paper started as an attempt to establish 
whether the local aspects of the dynamics of fields, which is 
apparent from the structure of the field equations, are  
fundamental in nature.  

In order to gain a deeper insight in the global aspects of the
dynamics of fields, a Lagrangian approach appears to be most 
suitable. In this approach, an action functional is constructed
from the field variables over
the entire spacetime. The dynamics of the field then follows 
by requiring that the action is extremal in the space of field
configurations. Establishing extremality of the action
amounts to showing that the action does not vary at first order,
for arbitrary infinitesimal perturbations of the field variables.
It is essential to note that the field perturbations which are
used to `test' the extremality of the action in the classical
description, are purely a mathematical construct. 
Further, we stress that the choice of the action functional is 
motivated with the  
aim to recover the field equations, and hence the Lagrangian
description has the same physical content as the field equation. 

At this point, let us formulate more precisely the question whether 
the local form
of the interactions in nature is fundamental.
On the one hand, it is well known that there exist conserved quantities 
which are  
related to global symmetries of the action \cite{noether} \cite{boek}.
Although the existence of conserved quantities suggests 
an underlying global aspect of the dynamics of classical
fields, this global aspect is in fact a consequence of applying Gauss's 
theorem to a four-divergence, which vanishes locally as
at each point in
our spacetime as a consequence of the field equation.
On the other hand, there is the question whether there can be
a non-local coupling between physical fields, which acts 
at the level of the field equation. In particular, one
would like to know whether non-local interactions at a 
classical level can emerge from an underlying quantum
theory for which the Lagrangian has the usual local
form. In this paper we will focus on this last question.

Let us now consider in some detail how 
classical field theory arises as a limit of an
underlying quantum field theory. 
According to the Feynman path-integral approach to quantum 
field theory, the expectation value of an operator $O$ which 
acts on a field $\psi$, is given by the formal expression,
\be\label{Z}
\ \langle O  \rangle = Z^{-1} \int \mb{d} [ \psi ]  O [ \psi ]
\ e^{\it{i} S [\psi] / \hbar },
\ee
where $S [\psi]$ is the action functional, $Z$ is a normalization constant, 
and $\mb{d} [ \psi ]$ is a measure on the space of field configurations 
(see, {\it e.g.}, \cite{path}).
The integral is evaluated over all field configurations (paths) 
which are  
continuous and which satisfy certain initial or periodicity 
conditions.
One should note that there is considerable difficulty involved in 
making the path-integral well defined, which is due to the fact 
that typical paths which contribute to the integral are 
non-differentiable. 
In our derivation, where we consider a free field, the different 
degrees of freedom decouple, and one can ignore those degrees 
of freedom which vary with infinite frequency.

As $\hbar$ approaches zero in expression (\ref{Z}), the oscillatory 
behavior of the integrand 
suggests that the integral is dominated by those 
field configurations $\psi$ which are in some sense near to a field 
configuration $\psi_0$ which extremizes the action.
Since $\hbar$ is close to zero when expressed in terms of macroscopic 
units of time and energy, one therefore 
expects that classical physics is accurately 
described by an action extremizing field configuration $\psi_0$.
The essential difference between this classical limit, 
and the classical theory which we discussed previously, is
the fact that in the former case there are physical field
perturbations which probe the phase-space nearby an
action extremizing configuration, while in the 
latter case these field perturbations are
purely a mathematical construct. 
As we will show in the following, this difference can  
give rise to an essentially different expression which describes the  
dynamics of the classical field.

As is well known, extremality of the action for $\psi_0$ 
implies that this configuration satisfies the classical field equations, 
{\em provided} that a surface term 
vanishes for all paths.
In a classical variational treatment, surface terms are set to zero trivially
by considering only paths which have compact support.
However, this restriction on the type of paths does
not occur in the sum over paths (\ref{Z}), and it seems natural 
to consider field configurations $\psi_0$ which extremize the 
action for the most general class of paths for which
extremality of the action can be defined.

Indeed, one should note that in a classical treatment of cosmological 
perturbations one does not  
normally assume that perturbations must have compact support.
However, if one accepts that classical perturbations do not have compact support, 
then it seems rather unnatural to require that quantum fluctuations about the 
classical field configurations have compact support. 
If this would be the case, then there would be a finite 
distance beyond which there are still classical perturbations 
while quantum fluctuations vanish. This appears to contradict
the Copernican principle, which is commonly adopted in 
cosmology. 

Considering the relation between classical and quantum physics,
it should be mentioned that the path-integral approach does not
only explain more than a classical approach 
({\it i.e.}, testable quantum effects),
but one also needs to assume more than in classical physics
({\it e.g.}, the existence of a classical regime \cite{coleman}, 
as well as various infinite 
subtractions \cite{path}).
One might therefore feel that the validity of the path-integral
approach is as questionable as the classical variational
approach, when it is applied to cosmological situations
where it has not been tested.  
When seen in this light, 
the classical assumption that field-perturbations  
are restricted to 
have compact support is not proven to be wrong, but rather, 
it represents one possible choice in
a more general class of boundary or asymptotic conditions.
Whichever point of view one favors, it seems interesting to
investigate the implications
of relaxing the assumption that field-perturbations 
must have compact support. We will discuss these 
implications in the following.

\section{Scalar field in FLRW geometry}

The line element of the FLRW geometry is
given by,
\be\label{FLRW}
\ \mb{d} s^2 = - \mb{d} t^2
\ + a^2 (t) \left[ \mb{d} \chi^2
\ +  c^{-2} \sinh^2 c \chi ( \mb{d} \theta^2 + \sin^2 \theta \mb{d} \phi^2 )
\ \right],
\ee
where $c = {\bf R}^+$ for the spatially open geometry, while the spatially
flat and closed geometry are obtained by taking the limit $c \downarrow 0$
or by choosing $c \in {\it i} \times {\bf R}^+$ respectively. 
We will refer to the
geometry with the line element (\ref{FLRW}) as ${\cal M}$, while a
spatial hypersurface of constant time $t$ is 
referred to as $\Sigma$.  

It follows directly from expression (\ref{FLRW}) that the surface of a
spatial sphere of constant radius $\chi_0$ grows as fast as the 
three-volume inside the sphere, when one considers the limit where 
$\chi_0 \rightarrow \infty$.
One may therefore expect that surface terms can be equally
important as volume terms when we take the infinite
volume limit in an open universe.
This situation is essentially different from the situation in
a spatially flat spacetime, where the surface of a spatial sphere
of constant radius $\chi$ grows by one power of $\chi$ less fast
then the three-volume  which is contained inside the sphere.

We will consider a scalar field $\psi$, which is described by
the Lagrangian density
\be\label{kl}
\ {\cal L} [ \psi ] = - \fr{1}{2} \sqrt{-g} 
\ (g^{\mu\nu} \partial_{\mu} \psi  \partial_{\nu} \psi
\ + m^2 \psi \psi ),
\ee
where $g_{\mu\nu}$ denotes the
FLRW metric (\ref{FLRW}), and $g = \det (g_{\mu\nu})$.

We define the action of the $\psi$-field as the integral
of the Lagrangian density (\ref{kl}) over the entire spacetime,
\be\label{S}
\ S [\psi ] := \int \mb{d}^4 x {\cal L} [\psi ].
\ee
Note that the integral in this expression does not need to converge.
This is not necessarily a problem if one is interested in calculating
the variation of the action under a change of the field from
$\psi$ to $\psi + \delta \psi$, where $\delta \psi$
is a suitably small `test-perturbation'.
The question arises which restriction one has to impose on
the test-perturbations $\delta \psi$ so that the first-order variation
$\delta S$ is well defined.
The first-order variation of the action (\ref{S}) follows by
the standard procedure of functional derivation,
\be\label{deltaS}
\ \delta S =
\ \int \mb{d}^4 x \left( \fr{\delta {\cal L}}{ \delta \psi } \delta \psi
\ + \fr{\delta {\cal L}}{  \delta \partial_{\mu} \psi } \delta \partial_{\mu} \psi \right).
\ee
By partially integrating equation (\ref{deltaS}), where the Lagrangian
is given by expression (\ref{kl}), we obtain
\be\label{deltaS2}
\ \delta S =
\ \int \mb{d}^4 x \sqrt{-g} \left[  \psi^{; \mu}_{;\mu}  - m^2 \psi  \right]  \delta \psi
\ - \int \mb{d}^4 x \sqrt{-g} ( \psi^{; \mu} \delta  \psi )_{ ; \mu } \;, 
\ee
where a semicolon denotes the covariant derivative.

Provided that the second term on the right-hand side of equation
(\ref{deltaS2}) vanishes for nonzero perturbations $\delta \psi$,
then the condition $\delta S =0$ implies the vanishing of the term
in brackets, and hence the field equation holds.
This is the case when we consider test-perturbations
$\delta \psi \in D$, where $D$ is defined as the class
of perturbations which are bounded and which have
compact support.
However, as we mentioned in the beginning of this section, the
restriction to test-perturbations $\delta \psi \in D$ does not
follow from known physical principles, when the spacetime
itself is non-compact.
Let us therefore try to determine the largest
class of test-perturbations for which the variation of the action is
well defined.
For a scalar field $\psi$, and a Lagrangian which is bi-linear
in the field variable, it is clear that square integrability of $\delta \psi$
is a necessary condition for the existence of the variation of the action (\ref
{deltaS2}), {\it i.e.}, we require $\delta \psi \in {L_2} ({\cal M})$.
It is not {\it a priori} clear whether $\delta \psi \in {L_2}({\cal M})$ 
is a sufficient
condition for the existence of the variation of the action
(\ref{deltaS2}), and it may be necessary to restrict the type
of test-perturbations further to ensure that $\delta S$ exists.
Assuming that we are able to determine the largest class of
test-perturbations $\delta \psi$ for which $\delta S$ exists,
then it remains a question whether there exist field configurations
$\psi_0$ such that $\delta S$ vanishes for all perturbations $\delta \psi$
about $\psi_0$.

Let us first address the question whether the restriction
$\delta \psi \in {L_2} ({\cal M})$ is sufficient to ensure the existence
of $\delta S$.
The answer to this question is negative, which we show by
an example where the contribution of surface terms to $\delta S$
diverges, while $\psi$ is a solution of the field equation and
$\delta \psi \in {L_2} ({\cal M})$.
Since we will focus on surface effects at {\em spatial} infinity, we
require that $\delta \psi$ can be square integrated over a spatial
hypersurface of constant time in the geometry (\ref{FLRW}),
{\it i.e.}, $\delta \psi \in {L_2} (\Sigma)$,
while we do not specify the time dependence of $\delta \psi$.
It is clear from the expression of the line element (\ref{FLRW})
that a square integrable test-perturbation $\delta \psi$
must approach zero faster than $1/\chi$ in the spatially flat
case, and faster than $e^{- \chi}$ in the spatially open case.
A specific example of a square integrable test-perturbation is
given by
\be\label{pert}
\ \delta \psi =  (1+ \chi)^{-(1 + \alpha)} \partial_\chi \psi 
\ \;\;\;\;\;  \mb{and}  \;\;\;\;\;
\ \delta \psi = e^{ - (1 + \alpha ) \chi} \partial_\chi \psi ,
\ee
in the spatially flat and open case respectively,
and $\alpha \in {\bf R}^+$.
By substituting expressions (\ref{pert}) for $\delta \psi$
into equation (\ref{deltaS2}), and using (\ref{FLRW}), we find
\be\label{c}
\ \delta S = -  4 \pi a^{-2} (t)
\   \int \mb{d} t \mb{d} \Omega \lim_{\chi \rightarrow \infty}  F (\chi)
\    (\partial_{\chi} \psi)^2,
\ee
where
$\mb{d}\Omega$ denotes the volume element on the unit
two-sphere, and $F (\chi) = \chi^{1 - \alpha}$ in the spatially flat
case, and $F (\chi) = e^{ ( 1 - \alpha ) \chi}$ in the spatially
open case.
Indeed, expression (\ref{c}) diverges for some values of
$\alpha \in (0,1]$, provided that the term $\partial_{\chi} \psi$
does not approach to zero as fast as $F^{-1/2} (\chi)$ in the
limit where $\chi \rightarrow \infty$.
The variation of the action (\ref{c}) can therefore be arbitrarily
large, for $\delta \psi \in L_2 (\Sigma)$.

Let us now address the question whether there exist
configurations of the $\psi$-field which extremize
the action for all $\delta \psi \in L_2 (\Sigma)$, in the
cosmologically interesting case where $\psi$
and $\partial_{\chi} \psi$ do not vanish at
spatial infinity. 
We show that the answer to this question is negative.
We will therefore use a result which is derived in the following, which
states that a field configuration which extremizes
the action for all $\delta \psi \in L_2 (\Sigma)$ must be
a solution of the field equation.
We combine this with the result which was derived earlier in this section,
which shows that a solution of the field equation for which
$\partial_{\chi} \psi$ does not approach to zero at spatial
infinity, does not extremize the action for all
$\delta \psi \in L_2 (\Sigma)$.
Hence, it follows that action extremizing configurations do not
exist for $\delta \psi \in L_2 (\Sigma)$ and $\partial_{\chi} \psi$
not approaching to zero at infinity.

In deriving the proof above, we assumed that a field configuration which extremizes 
the action for all $\delta \psi \in L_2 (\Sigma)$ must be a solution of the field 
equation.
In order to proof this, let us recall that for $\delta \psi \in D$, {\it i.e.}, 
the class of test-perturbations which are bounded and which have compact 
support, extremality of the action implies that the field equation
holds and vice-versa.
Configurations which do not satisfy the field equation can therefore not
extremize the action for all $\delta \psi \in D$,
and since $D \subset L_2 (\Sigma)$ these configurations do not extremize the
action for all $\delta \psi \in L_2 (\Sigma)$. Hence it follows that
a field configuration which extremizes 
the action for all $\delta \psi \in L_2 (\Sigma)$ must be
a solution of the field equation, which proves our assumption.

The observation that action extremizing configurations do not in general
exist for $\delta \psi \in L_2 (\Sigma)$, implies that
the usual identification between classical physics and
action extremizing configurations becomes ambiguous  
when we allow for perturbations which do not fall
off sufficiently fast at infinity.
There are several ways by which one could try to resolve
the problem which is posed by the non-existence of 
extremal action configurations for test-perturbations
$\delta \psi \in L_2 (\Sigma)$.
We will discuss these possible solutions in the following. 

First, let us recall that the restriction $\delta \psi \in L_2 (\Sigma)$
was found to be {\em necessary} to ensure finiteness of $\delta S$, but 
due to the contribution of a surface term to $\delta S$ 
this restriction is not {\em sufficient}.
This observation suggests that the class of test-perturbations
$\delta \psi$ should be restricted
further, such that $\delta S$ is finite for all $\delta \psi$.
Although finiteness of $\delta S$ is easily achieved by requiring  
that the test-perturbations $\delta \psi$ fall off sufficiently 
fast, this does not imply that extremal action configurations exist
in the space of test-perturbations for which 
$\delta S$ is finite.
The reason for this is that the existence of extremal 
action configurations requires that the surface term 
contribution to $\delta S$ vanishes completely, 
which is clearly a stronger restriction on 
$\delta \psi$ than the condition that $\delta S$ is finite.  
Although one could restrict $\delta \psi$ to ensure that the surface term 
contribution to $\delta S$ vanishes completely, this would be rather
add-hoc since it is not shown that this is the only possible
restriction on the class of test-perturbations for which 
extremal action configurations exist.

Instead of restricting the class of test-perturbations, one could also 
attempt to remove the contribution of surface terms to $\delta S$ 
by modifying the Lagrangian density (\ref{kl}).  
Let us therefore note that the choice of the Lagrangian density
is motivated by the fact that one recovers the Klein-Gordon equation,
provided that the variation of the action
and the surface term in equation (\ref{deltaS2}) vanish.
In the classical variational approach, where surface terms are
made to vanish by assuming boundary conditions on $\delta \psi$,
one therefore has the freedom to
add a term to the Lagrangian density
which has the form of a four-divergence,
since the variation of this term equals  
a vanishing surface term.
In this section we questioned the assumption  
that the surface term in equation (\ref{deltaS2}) vanishes in perturbed
flat and open FLRW spacetimes.
However, it is conceivable that one can add a
four-divergence term to the Lagrangian (\ref{kl})
such that its variation cancels the surface term
in equation (\ref{deltaS2}).
Indeed, in the context of Hamiltonian cosmology, 
as well as in quantum cosmology, it 
appears to be natural to add a surface term to the 
Einstein-Hilbert action which has the property that its 
variation cancels an 
identical term which arises from the variation of  
the Einstein-Hilbert action \cite{regge} - \cite{hg2}.

Let us now consider whether the same possibility  
exists in the case where we are dealing with a scalar field. 
We therefore add a generic surface term to the action (\ref{S}), 
which has the form
\be\label{calb}
\ S_B [ \psi ] =  \fr{1}{2} \int \mb{d}^4 x \sqrt{-g}  B^{\mu}_{;\mu} \;,
\ee
where $B^{\mu} = B^{\mu} [\psi]$, and then we consider whether
the variation of this surface term may cancel the surface term
in equation (\ref{deltaS2}).
The variation of $B^{\mu}$ 
follows by the method of functional derivation, {\it i.e.}, 
treating $\psi$ and $\partial_{\nu} \psi$ as 
independent variables:
\be\label{nnx}
\ \delta B^{\mu} =  \fr{\delta B^{\mu}}{\delta \psi} \delta \psi
\ + \fr{\delta B^{\mu}}{\delta \partial_{\nu} \psi} \delta \partial_{\nu} \psi , 
\ee
where we used that $B^{\mu}$ cannot depend on 
higher than first-order derivatives of $\psi$.
It is clear that any dependence of $B^{\mu}$ 
on higher than first-order derivatives of $\psi$
contributes terms to the variation of the action
which are proportional to the variation of higher
than first-order derivatives of $\delta \psi$.
These terms cannot cancel against the surface term in equation
(\ref{deltaS2}), which contains at most first-order derivatives
of $\delta \psi$,
although a cancellation was required.

The requirement that the surface term in equation (\ref{deltaS2})
cancels the surface term which arises from the
variation of $S_B$ results in the conditions 
\be\label{cz1}
\ \fr{\delta B^{\mu}}{\delta \psi} =  \psi^{;\mu}, \;\;\;\;\; \mb{and} \;\;\;\;\;
\ \fr{\delta B^{\mu}}{\delta \partial_{\nu} \psi} = 0,
\ee
for all $\mu,\nu$, and we used expression (\ref{nnx}).
The first condition in equation (\ref{cz1}) constrains $B^{\mu}$ to
be of the form $B^{\mu} = \psi^{;\mu} \psi +$ c$_1$,
where $c_1$ is a functional which does not depend on $\psi$,
while the second condition constrains $B^{\mu}$ to be
a functional which does not depend on $\partial_{\mu} \psi$.
Clearly, both requirements are exclusive, 
and there exists no
functional $B^{\mu}$ such that the variation of $S_B$, (\ref{calb}), 
cancels the surface term in equation (\ref{deltaS2}).
Note, however, that the precise form of the surface term in
equation (\ref{deltaS2}) does change by adding a term of the
form (\ref{calb}) to the action.
Hence, the contribution of a surface term to the variation of the
scalar field action (\ref{S}) appears to be generic, although its
precise form is ambiguous. 
In the following calculation we will retain the surface term
which appears in equation (\ref{deltaS2}), which means that 
we assume $S_B$ to vanish. 

Having considered the possibility to adopt further restrictions on
the type of test-perturbations, as well as modifying the action
by adding a surface term contribution, we have not found an
argument which shows us that we can neglect the 
contribution of a surface term to the variation of the action.
However, taking the surface term in equation (\ref{deltaS2})
seriously confronts us with the
problem that field configurations which extremize the 
action in the space of test-perturbations for which
$\delta S$ is well defined, do not in general exist.  
It should be noted, however, that 
the non-existence of action extremizing field configurations
does not need to be a problem if one could show that 
those test-perturbations
for which the action functional is not extremal, have a zero
phase-space measure in the space of fields $\psi$. 
Indeed, it is clear that paths of the form (\ref{c}),
which yield large surface terms at spatial infinity, are highly
special in the sense that the asymptotic behavior of these paths
is correlated with the field $\psi$ about which we expand.
Therefore, one expects that these paths occupy a very small
amount of phase-space in the space of field configurations
in which extremality of the action is considered, and
their relevance for the dynamics of the $\psi$-field may be
negligible.
Note, however, that precisely the same argument applies
to the case where $\delta \psi \in D$, since in this case
$\delta \psi$ is specified to be exactly equal to zero for
arbitrarily large radii $\chi$.
In order to make these considerations quantitative, 
it is necessary to introduce a measure on the phase-space of the
$\psi$-field. 
We will address this problem in the following sections.


\section{Perturbations in open FLRW}\label{ead}

In order to obtain a quantitative description of the space of
field configurations of
the scalar field $\psi$, it is useful to decompose $\psi$
and test-perturbations $\delta \psi$ in terms of eigenfunctions
of the spatial Laplacian which are complete in the space $L_2$ of
functions which are square integrable on the hypersurfaces 
$\Sigma (t)$. The reason why it is convenient to use
eigenfunctions of the spatial Laplacian, is that this operator 
is present in the expression for the variation of 
the action (\ref{deltaS2}).
When we ignore the surface term, it is therefore clear that 
each eigenfunction only couples to itself, and the dynamics of each mode
is independent of the dynamics of all other modes. 

Let $Q (x)$ be a solution of 
the Helmholtz equation, {\it i.e.},
\be\label{helm}
\ Q^{; i}_{; i} +  ( k / a)^2 Q = 0,
\ee
where $; i$ denotes the covariant derivative with respect to the coordinate
$x^{i} \in \{ r,\theta,\phi \}$ in the geometry (\ref{FLRW}),
$a=a(t)$ denotes the scale factor, and $k \in {\bf R}^+$.
In the following, we concentrate on the spatially open geometry
(\ref{FLRW}), while we consider the spatially flat spacetime as
a limiting case of the spatially open geometry.
A basis of solutions of equation (\ref{helm}), which are
complete in the space of $L_2$ functions on $\Sigma (t)$, 
and which factorize
in terms of an angular and a radially dependent part,
is given by
\be\label{Zqlm}
\ Z_{q l m} = \Pi_{q l} (\chi)
\ Y_{l m} (\theta,\phi) ,
\ee
where $Y_{l m}$ are the standard spherical harmonics on the unit two-sphere,
and the radially dependent functions $\Pi_{q l} (\chi)$ are solutions of the
equation
\be\label{fr}
\ \fr{1}{g_2 } \fr{\p}{\p \chi} {g_2} \fr{\p}{\p \chi}  \Pi_{q l}  (\chi ) =
\  \left(  k^2  - \fr{l ( l + 1 )}{g_2} \right)  \Pi_{q l} ( \chi),
\ee
where $g_2 = c^{-2} \sinh^2 c \chi$.
Equation (\ref{fr}) has solutions of the form,
\be\label{gr}
\ \Pi_{q l} (\chi) = N_{q l}  (\sinh c \chi )^l
\ \left( \fr{-1}{ \sinh c \chi} \fr{\mb{d}}{\mb{d} \chi} \right)^{l + 1}
\ \cos (q c \chi),
\ee
where $q$ is defined by $q^2 =  k^2 / c^2 -1$, and  
\be
\ N_{q l} := \sqrt{\fr{2}{\pi}}
\  \left[  \prod_{n=0}^l (n^2 + q^2)
\ \right]^{-1/2}
\ee
is a normalization factor
\cite{fabri, garcia}.
Notice that the $q =0$ mode 
solves the Helmholtz equation (\ref{helm})
with a {\em nonzero} eigenvalue equal to
$-c^2/a^2$, which equals $\fr{1}{6}$ times the 
spatial curvature in the geometry (\ref{FLRW}).  

The radial solutions for the spatially flat geometry are
obtained by taking the limit $c \downarrow 0$ in
expression (\ref{gr}), keeping $k$ fixed,
\be
\ \lim_{c \downarrow 0}   \Pi_{q l} (\chi)
\   = \sqrt{\fr{2}{\pi}} k j_l ( k \chi),
\ee
where $j_l$ denotes the spherical Bessel function \cite{abbott}.
From now on, we assume that the spacetime is open, such that
$c \in {\bf R}^+$, and without loss of generality we may set 
$c=1$ in expression (\ref{FLRW}) by absorbing a 
factor $c$ in the definition 
of the comoving radial coordinate $\chi$ and by absorbing
a factor $c^{-1}$ in the
definition of the scale factor $a(t)$.

It follows from expression (\ref{gr}) that the radial functions $\Pi_{q l}$ 
can be written as the product of an oscillating factor $\cos q \chi$ or
$\sin q \chi$, and a factor which approaches to zero exponentially
as $\sinh^{-1} \chi$ in the limit where $\chi \rightarrow \infty$.
Since the modes $Z_{q l m}$ with $q \in {\bf R}^+$ vary
at comoving length scales which are typically smaller than the
curvature scale which we have set equal to one
in the FLRW geometry (\ref{FLRW}),
these modes are called {\em subcurvature modes}.

There exist solutions of the Helmholtz equation (\ref{helm}) for which
$k^2 \in (0,1]$, which corresponds to imaginary values of
$q \in {\it i} \times (0,1]$.
The explicit expression for these modes is still given by equation
(\ref{gr}), where the factor $\cos (q \chi )$ is replaced by
$\cosh ( |q| \chi )$.
The modes $Z_{q l m}$ with $q \in {\it i} \times (0,1]$ approach to zero
as a constant times $\exp ((|q| -1) \chi )$ in the limit where
$\chi \rightarrow \infty$, and since they vary at length
scales greater than the curvature scale one calls them 
{\em supercurvature modes}.

We define the spatial integration operation by
\be\label{int-inf}
\  \langle  f \rangle := \lim_{\epsilon \downarrow 0}
\  \langle  f \rangle (\epsilon)
\ee
where
\be\label{int-fin}
\ \langle  f \rangle (\epsilon) :=
\  \int \mb{d} \Omega \int_0^{1/ \epsilon} \mb{d} \chi \sinh^2 (\chi)  f .
\ee
and $\mb{d} \Omega^2$ denotes the volume element on the unit two-sphere.
The subcurvature modes $Z_{q l m} (q \in {\bf R}^+)$ are orthonormal
with respect to spatial integration,
\be\label{norm}
\ \langle Z_{q l m}  Z_{q^{\prime} l^{\prime} m^{\prime}} \rangle
\ = \delta (q - q^{\prime}) \delta_{l l^{\prime}} \delta_{m m^{\prime}},
\ee
and they are known to be complete in the space $L_2 (\Sigma)$ \cite{bander},
which consists of equivalence classes of functions $f$ for which 
$\langle |f|^2 \rangle$ exists, where 
we identify functions $f$ which differ only on a set
of Lebesgue measure zero.

For the supercurvature modes, the indefinite integral over the radius
in expression (\ref{norm}) does not exist, so that these modes cannot
be normalized in the $L_2 (\Sigma )$ sense.
Furthermore, expression (\ref{norm}) diverges when only
one of the modes $Z$ corresponds to a supercurvature mode, and
$l = l^{\prime}$ and $m=m^{\prime}$.
Therefore, the supercurvature modes cannot be decomposed in terms
of the subcurvature modes.
Mechanisms which may be responsible for the generation of supercurvature
perturbations in open spacetimes
have been investigated in \cite{bharat,kaiser}.

The $\psi$-field may be expanded in terms of the modes $Z_{q l m}$,
\be\label{ge}
\ \psi (  x , t)  = \psi^{-} (  x , t)
\ + \psi^{+} (  x , t),
\ee
where
\be\label{psisub}
\ \psi^{-} (x,t) := \sum_{l m}  \int_0^{\infty} \mb{d} q
\ \psi_{q l m} (t) Z_{q l m} (  x ),
\ee
\be\label{psisuper}
\ \psi^{+} (x,t) := \sum_{l m} \int_{0}^{\it{i}} \mb{d} \bar{q}
\  \psi_{\bar{q} l m} (t) Z_{\bar{q} l m} ( x  ),
\ee
where $ x  = \{ \chi , \theta, \phi \}$, and the integration
over $\bar{q}$ runs along the imaginary
axis in the complex $\bar{q}$-plane.

An important class of perturbations, which is believed to occur in
the early universe, corresponds to the case where the coefficient 
of each independent mode 
is chosen according to a Gaussian probability distribution
(see, {\it e.g.}, \cite{ran1} - \cite{ran3}).
For this type of perturbation, which is called a `Gaussian perturbation'
or `random-field',
there are no correlations between the
coefficients $\psi_{q l m}$ for different values of $q,l$, and $m$. 
The statistical properties of a random-field 
are determined by the variance of the Gaussian probability
distribution, which we call $\sigma$.
In the generic case, where $\sigma$ depends on $q,l,$ and $m$,
one cannot determine the variances $\sigma (q,l,m)$ from a single realization
of a random-field, which is determined by the set of coefficients 
$\psi_{q l m}$.
Instead, one would need an infinite {\em ensemble} of random-fields,
in order to deduce the statistical properties, {\it i.e.},
the variances $\sigma (q,l,m)$, according to which these
random-fields are generated.

Let us now define the {\em ensemble average} of a functional as the
weighted sum of this functional over all random-fields
in an ensemble, where the weight factor is given by the
probability for each specific random-field to occur.
This allows us to define the two-point correlation function
of the $\psi$-field
as the ensemble average of $\psi (x)$ times $\psi (x^{\prime})$.
A random-field $\psi (x)$ is said to be statistically homogeneous
and isotropic when the two-point correlation function
is invariant under the group of isometries on $\Sigma$,
{\it i.e.}, the group of rotations and spatial translations.
Clearly, the two-point correlation function of a statistically homogeneous
and isotropic random field can only be a function of a distance measure which
is invariant under the group of isometries on $\Sigma$, 
and we can take this distance measure
to be the length $d(x,x^{\prime})$ of a geodesic which relates the
points $x$ and $x^{\prime}$.
In can be shown that statistical homogeneity and isotropy of a
random-field $\psi (x)$ holds if and only if the variances $\sigma$
do not depend on the labels $l$ and $m$ \cite{yag}.

Although it seems rather artificial to introduce the concept of an ensemble
in the context of cosmology, since we can only observe one universe,
a physical interpretation of the ensemble average is provided by
the property of {\em ergodicity}. In the context of random-fields, ergodicity is
defined as the equivalence of ensemble averaging and spatial averaging,
where the spatial average of the two-point correlation function
is defined by summing $\psi (x)$ times $\psi (x^{\prime})$ over
random sets of points $x$ and $x^{\prime}$ for which the
geodesic distance $d(x,x^{\prime})$ has a specific value.
In the case where $\Sigma$ is a Euclidean three-space,
ergodicity can be proven to hold under fairly weak
assumptions \cite{ran1}, but for a hyperbolic three-space
no proof seems to be known, while it is usually assumed.

In the following, we will assume a Gaussian statistically homogeneous
and isotropic spectrum of subcurvature perturbations.
One should note that this type of perturbation 
cannot be square integrated. 
This follows by substituting the expansion of $\psi$, (\ref{psisub}),
into the hypersurface integral (\ref{int-inf}) and using the orthonormality
relation (\ref{norm}). 
The resulting expression contains an indefinite sum over $l$ and $m$ 
of the squared coefficient $\psi_{q l m}$, and this sum diverges when
the variance $\sigma (q)$ is nonzero. 
It is therefore clear that the property of non-square integrability is not
specifically related to the presence of supercurvature modes.

\section{Extremal action dynamics}\label{dyn}

Let us now calculate the variation of the action (\ref{deltaS2}),
which is evaluated over a bounded spatial volume $V (\chi_0)$,
which we define as those points in the geometry (\ref{FLRW})
for which $\chi < \chi_0$, 
and then we consider the limit where $\chi_0\rightarrow \infty$.
We obtain
\[
\ \delta S = \int \mb{d} t   \lim_{\chi_0 \rightarrow \infty}
\  \left[   a^3 \int \mb{d} \Omega^2 \int_0^{\chi_0} \mb{d} \chi
\ \sinh^2 \chi
\ \delta \psi   \left( \fr{1}{\sqrt{-g}} \partial_{\mu} g^{\mu\nu} \sqrt{-g}
\ \partial_{\nu}
\ -  m^2 \right)  \psi
\ \right.
\]
\be\label{ccc}
\ -    a \left. \left. \sinh^2 \chi \int \mb{d} \Omega^2
\   \delta \psi  \p_{\chi} \psi  \right|_{\chi= \chi_0} \right],
\ee
where $a = a(t)$.
Using the definition of the integration operation (\ref{norm}),
expression (\ref{ccc}) can be written in the form,
\[
\ \delta S = \int \mb{d} t \left[ a^3
\ \left\langle \delta \psi
\ \left( \fr{1}{\sqrt{-g}} \partial_{\mu} g^{\mu\nu} \sqrt{-g} \partial_{\nu} - m^2 \right)
\ \psi \right\rangle \right.
\]
\be\label{kk}
\ - \left. \left. a \lim_{\chi_0 \rightarrow \infty}
\   \sinh^2 \chi_0 \int \mb{d}  \Omega \delta \psi  \p_{\chi} \psi 
\  \right|_{\chi = \chi_0} \right].
\ee
We will consider separately the cases where the expansion of the
field $\psi$ includes only subcurvature modes, and the 
case where the expansion includes supercurvature modes
as well.

\subsection{Open spacetime with subcurvature perturbations}\label{osub}

Let us first consider the case where the field $\psi$ can be
expanded in terms of only subcurvature modes, {\it i.e.},
we assume that $\psi_{q l m} = 0$ for all $q \in \it{i}\times (0,1]$,
so that only the first term in the expansion of the field
(\ref{ge}) is nonzero.
Equation (\ref{kk}) can then be evaluated separately for each
mode, by substituting the expansion (\ref{ge}) into expression
(\ref{kk}), and using the orthonormality relation (\ref{norm}).
We obtain
\be\label{vvd}
\ \delta S = \int \mb{d} t \int \mb{d} q \sum_{l,m}
\  \delta \psi_{q l m} (t) \left[ a^3
\ \left( \fr{1}{\sqrt{-g}} \partial_0 g^{00} \sqrt{-g} \partial_0
\ - a^{-2} (t) k^2  - m^2 \right)    \psi_{q l m} (t) \right.
\ee
\[
\ \left. \left. - \lim_{\chi_0 \rightarrow \infty}  a \sinh^2 \chi_0
\ \int \mb{d} q^{\prime}  \psi_{q^{\prime} l m}
\  \Pi_{q l} \partial_\chi  \Pi_{q^{\prime} l} \right|_{\chi=\chi_0} \right].
\]
The requirement that the variation of the action vanishes for nonzero
perturbations $\delta \psi_{q l m} (t)$ implies an equation
of motion for each perturbation component $\psi_{q l m} (t)$, namely,
\be\label{FJ}
\ \left( \fr{1}{\sqrt{-g}} \partial_0 g^{00} \sqrt{-g} \partial_0
\ - a^{-2} k^2  - m^2  \right)    \psi_{q l m} (t) =  J_{q l m},
\ee
where
\be\label{J}
\ J_{q l m}  :=  \lim_{\chi_0 \rightarrow \infty}
\ \left[ \left. a^{-2} \sinh^2 \chi_0 \int \mb{d} q^{\prime} \psi_{q^{\prime} l m}
\  \Pi_{q l} \partial_\chi  \Pi_{q^{\prime} l}
\ \right|_{\chi =\chi_0} \right].
\ee
Note that $J_{q l m}$ acts as a {\em source term} in equation (\ref{FJ}),
and this term couples perturbations which have the same angular
wave numbers $l$ and $m$.
One would like to know whether the limit in expression (\ref{J}) exists,
and whether or not this term can be neglected.
In order to answer this question, we need to evaluate the integral
over $q^{\prime}$ of the distribution $\psi_{q^{\prime} l m}$, which
is multiplied by a factor which is of order unity.
According to equation (\ref{norm}) and  (\ref{ge}), the distribution
$\psi_{q^{\prime} l m}$ can be defined by,
\be\label{psik}
\ \psi_{q^{\prime} l m}  =  \lim_{\epsilon \downarrow 0}
\ \psi_{q^{\prime} l m} (\epsilon),
\ee
where
\be\label{psie}
\ \psi_{q^{\prime} l m} (\epsilon) :=
\ \langle Z_{q^{\prime} l m} \psi \rangle (\epsilon)
\ee
and the limit $\epsilon \downarrow 0$ should be evaluated
after the integration over $q^{\prime}$ is performed.
When we integrate over a bounded volume, then the modes
$Z_{q l m}$ are dependent in the sense that their
overlap
$\langle Z_{q l m} Z_{q^{\prime} l m} \rangle (\epsilon)$
is nonzero and of the order of $\epsilon^{-1}$ for $q - q^{\prime}$
of the order of $\epsilon$.
The number of independent modes in a fixed $q^{\prime}$-interval
therefore tends to diverge as $\epsilon^{-1}$ in the limit where
$\epsilon \downarrow 0$.
In the previous section, we introduced the concept of a Gaussian perturbation.
In order to generate a Gaussian perturbation which has an amplitude of order one,
the coefficients $\psi_{q l m}$  
in the expansion of the field (\ref{psisub}) need to be  
uncorrelated for values of $q$ differing more than $\epsilon$, 
while the amplitude of the coefficients must diverge as 
$\epsilon^{-\fr{1}{2}}$ when 
$\epsilon \downarrow 0$. 
The asymptotic behavior of the integral over $q^{\prime}$ in 
expression (\ref{J}) can therefore
be estimated as the sum of $\epsilon^{-1}$ uncorrelated numbers
which are of the order of $\epsilon^{-1/2}$, multiplied by
a $q^{\prime}$-interval which is of the order of $\epsilon$.
In the limit where $\epsilon \downarrow 0$, the term between brackets
in expression (\ref{J}) will therefore remain of order one, and the
expression does not converge.
Note, however, that the left-hand side of the equation of motion
(\ref{FJ}) is proportional to the coefficient $\psi_{q l m}$,
which diverges as $\epsilon^{-1/2}$ in the limit where
$\epsilon \downarrow 0$.
We therefore find that the source term on the right-hand side 
of equation (\ref{FJ})
can be neglected in the infinite volume limit, when the perturbations
of the field are Gaussian and of the subcurvature type.


\subsection{Open spacetime with supercurvature perturbations}\label{osup}

Let us now attempt to derive an equation of motion for the
$\psi$-field, in the
case where the expansion of the $\psi$-field (\ref{ge}) includes
supercurvature perturbations.

We may therefore substitute the expansion of the $\psi$-field (\ref{ge})
in the expression for the variation of the action (\ref{ccc}),
which yields,
\be\label{ccc2}
\ \delta S = \int \mb{d} t
\ \lim_{\chi_0 \rightarrow \infty}
\ee
\[
\ \times  \left[ a ^3  \int  \mb{d} \Omega^2 \int_0^{\chi_0} \mb{d} \chi
\ \sinh^2 \chi
\ \delta \psi  \left( \fr{1}{\sqrt{-g}} 
\ \partial_{\mu} g^{\mu\nu} \sqrt{-g}
\ \partial_{\nu}
\  - m^2 \right) ( \psi^- + \psi^{+} ) \right.
\]
\[
\ -  a \left. \left. \int \mb{d} \Omega^2   \sinh^2 \chi
\  \delta \psi  \p_{\chi} (\psi^{-} + \psi^{+} )
\ \right|_{\chi= \chi_0} \right].
\]
Using the definition of the integration operation (\ref{int-inf}),
and expression (\ref{ge}), we recover expression (\ref{FJ}), 
with an additional source term which accounts for the 
coupling between subcurvature and supercurvature 
perturbations, {\it i.e.},
\be\label{FJ2}
\ \left( \fr{1}{\sqrt{-g}} \partial_0 g^{00} \sqrt{-g} \partial_0
\ - a^{-2} k^2  - m^2  \right)    \psi^-_{q l m} (t) = 
\ J_{q l m} + J^{+}_{q l m},
\ee
where $q \in {\bf R}^+$, $J_{q l m}$ is given by expression (\ref{J}), and
\[
\ J^{+}_{q l m}  :=  \lim_{\chi_0 \rightarrow \infty}
\  \int_0^{\it{i}} \mb{d} \bar{q} \left[
\ \left( \fr{1}{\sqrt{-g}} \partial_0 g^{00} \sqrt{-g} \partial_0
\ - a^{-2} k^2  - m^2
\  \right) \psi^{+}_{\bar{q} l m} (t)  \right.
\]

\be\label{J2}
\ \times \left.  \left. \int^{\chi_0}_0 \mb{d} \chi \sinh^2 \chi \Pi_{q l} \Pi_{\bar{q} l}
\ +  a^{-2} \sinh^2 \chi_0 
\ \psi^{+}_{\bar{q} l m}
\ \Pi_{q l} \partial_{\chi} \Pi_{\bar{q} l}  \right|_{\chi=\chi_0}
\ \right].
\ee
Note that both terms which contribute to expression (\ref{J2})
diverge exponentially in the limit where $\chi_0 \rightarrow \infty$, 
and the limit in this expression does not
exist, unless the divergent terms cancel. 
Let us therefore observe that the two terms at the right-hand side of equation 
(\ref{J2}) diverge exponentially as $\exp |\bar{q} \chi |$, 
(see section \ref{ead}),
and both terms oscillate due to the radial function $\Pi_{q l m}$.
A cancellation of the divergent terms in equation (\ref{J2}) 
requires that both terms oscillate with the same phase.
By re-writing equation (\ref{J2}), using,
\be\label{mm}
\ \int_0^{\chi_0} \mb{d} \chi \sinh \chi
\ \Pi_{q l} \Pi_{\bar{q} l} =  \fr{\sinh^2 \chi_0}{q^2 - \bar{q}^2  } \left|
\ \Pi_{q l} \partial_{\chi}  \Pi_{\bar{q} l}
\ - \Pi_{\bar{q} l}  \partial_{\chi}  \Pi_{q l}
\  \right|_{\chi = \chi_0},
\ee
one finds that $J^{+}_{q l m}$ 
diverges as the product of an exponential factor 
$\exp (|\bar{q}| + 1) \chi $, multiplied by the sum of two terms
which oscillate out of phase as  
$\Pi_{q l}$  and $\partial_{\chi}  \Pi_{q l}$, 
respectively.
Therefore, the right-hand side of equation (\ref{FJ2}) diverges, 
and we cannot use this equation to describe the time-evolution 
of the perturbation component $\psi_{q l m} (t)$. 
Recall that in the absence of supercurvature perturbations,  
surface terms appeared to give rise to a negligible correction 
to the equation of motion for each perturbation component 
$\psi_{q l m} (t)$, which followed by requiring that
$\delta S =0$ for all $\delta \psi \in L_2 (\Sigma)$.
When supercurvature perturbations are present, equations (\ref{ccc2})
and (\ref{FJ2}) show that it is precisely a surface term which 
contributes a divergent term to the variation of the action for all 
$\delta \psi \propto Z_{q l m}$.
In this case, the extremal action condition $\delta S=0$ cannot be
satisfied for all $\delta \psi \in L_2 (\Sigma )$, irrespectively of 
the equation of motion which the field satisfies.  
It is however clear that the condition $\delta S=0$ must have 
solutions when test-perturbations are confined to
some subspace of $L_2 (\Sigma)$  
for which $\delta S$ is well defined.
We will determine these subspaces in the following.

According to expressions (\ref{kk}) and (\ref{ge}), the surface term
which contributes to $\delta S$ behaves asymptotically as
$\delta \psi$ times  a factor $\sinh^2 \chi \partial_{\chi} \psi^+$ in the
limit where $\chi \rightarrow \infty$.
The contribution of surface terms to the variation of
the action (\ref{ccc}) will therefore be finite and convergent,
provided that 
$ \sinh^2 \chi \delta \psi \partial_{\chi} \psi^+$
converges when
$\chi \rightarrow \infty$.
Let us now {\em define} the class of test-perturbations $\{ \delta \psi \}_c$
by the requirement that $ \sinh^2 \chi \delta \psi \partial_{\chi} \psi^+$
converges to a constant $c \in {\bf R}$ when
$\chi \rightarrow \infty$.

Note that it follows from the definition of $\{ \delta \psi \}_c$ that
$\{ \delta \psi \}_c$ contains $D$, i.e., the class of functions
which are bounded and which have compact support.
As is well known, the class of functions $D$ is infinite dimensional
in the sense that there exists a denumerable infinite set of linearly
independent basis-functions which is complete in $D$
\cite{friedlander}, and therefore 
$\{ \delta \psi \}_c$ must be infinite dimensional, 
for arbitrary $c \in {\bf R}$.
It is therefore not clear whether one class
of test-perturbations $\{ \delta \psi \}_c$ for 
some specific value of $c \in {\bf R}$ dominates
in terms of the phase-space which
is occupied by these 
test-perturbations.
We will make this statement more precise in the following
section, where it is shown that that the classes of
test-perturbations $\{ \delta \psi \}_c$, for different
values of $c\in{\bf R}$,
are equivalent up to variations
with vanishing $L_2 ({\cal M})$-norm.

Summarizing, we found that the contribution of
surface terms to the variation of the action diverges 
for square integrable field
perturbations which do not 
fall off at a specific rate, depending on the
spectrum of supercurvature perturbations.
In the presence of supercurvature perturbations,
extremality of the action can therefore 
only be defined with respect to 
a restricted class of field perturbations. 
Surface terms contribute a non-trivial source term to 
the standard Klein-Gordon equation,
but the magnitude thereof depends on the
choice of the restricted class of
test-perturbation with respect to
which the action is extremized.
The dynamics of the `classical' field configurations 
therefore
remains undetermined, unless one finds
a physical argument which constrains 
the phase-space of the $\psi$-field
uniquely.

\section{Quantum correlations}\label{q}

In the previous section we showed that surface terms  
constrain the phase-space of test-perturbations
for which the variation of the Klein-Gordon
action is finite,
in an open FLRW spacetime with supercurvature perturbations. 
One may also question whether the nontrivial surface terms which we
found have an effect
on quantum correlations of the $\psi$-field.
As is clear from expression (\ref{Z}), the quantum correlation function
of the $\psi$-field can be expressed as a weighted integral over all
continuous field configurations, and the weight factor
depends on the source term $J^+$, which may be infinite.

The two-point correlation function is given by
the formal expression (see, {\it e.g.}, \cite{path})
\be\label{tau}
\  \tau (x,x^{\prime}) :=  Z^{-1} \int \mb{d} [ \psi ]
\  \psi (x)  \psi (x^{\prime})
\ e^{ \it{i} S [\psi ] / \hbar  },
\ee
where $x$ denotes the set of
coordinates on ${\cal M}$. 

The standard method to calculate the two-point correlation function
is to expand the field $\psi$, about some background 
configuration $\psi_0$, in terms of a denumerable
complete set of solutions of 
the four-dimensional Helmholtz equation 
(see, {\it e.g.}, \cite{sh} for the details involved in this 
calculation).
Since $L_2 ({\cal M})$ is known to be separable, there exists a denumerable
and complete set of solutions, which we call $\psi_i$, and
we can choose these solutions to be orthonormal in $L_2 ({\cal M})$.
A generic expansion of the field $\psi$, about  
a configuration $\psi_0$, takes the form 
\be\label{owl}
\ \delta \psi := \psi - \psi_0 = \sum_i a_i \psi_i,
\ee
where $a_i \in {\bf R}$.
Further, the measure on the space of the 
field $\psi$ can be expressed 
in terms of the coefficients $a_i$, {\it i.e.},
\be\label{measure}
\ \mb{d} [\psi] = \prod_{i} \mu \mb{d} a_i,
\ee
where $\mu$ is a normalization constant with the dimension 
of inverse length, and the indefinite product runs over all
values of the label $i$.

By substituting the expansions of the field (\ref{owl}) and 
the measure (\ref{measure}) into the expression for
the correlation function (\ref{tau}), the path-integral
can be evaluated explicitly. 
Assuming that there are no nontrivial source terms of the
kind which we discussed in the previous section, 
then the standard expression for the two-point
correlation function follows in terms of the complete
set of modes $\psi_i$. 
We will not repeat this calculation here, which can be found,
{\it e.g.}, in \cite{sh}, but instead
we will consider what is the effect on
the two-point correlation
function (\ref{tau}) when there is a nontrivial 
source term $J^+ [\psi]$ which contributes to
the variation of the action.

Let us now define the set of functions
$\tilde{\psi}_i \in \{ \delta \psi \}_0$,  
which satisfy the property that the linear span
of the modes $\tilde{\psi}_i$ is dense 
in $\{ \delta \psi \}_0$,
and the modes $\tilde{\psi}_i$ are chosen 
so that they are orthonormal with respect to the 
$L_2 ({\cal M})$-inner product.
We would like to show that the modes $\tilde{\psi}_i$ are complete
in $L_2 ({\cal M})$.
Note that the class of functions $D ({\cal M})$, which are bounded and 
which have
compact support on ${\cal M}$, is contained in $\{ \delta \psi \}_0$.
But $D ({\cal M})$ is known to be dense in 
$L_2 ({\cal M})$ with the $L_2 ({\cal M})$-norm, and therefore
the linear span of the modes 
$\tilde{\psi}_i$ must be dense in $L_2 ({\cal M})$.
At this point, let us note that the set of functions
$L_2 ({\cal M})$,
with the 
$L_2 ({\cal M})$-inner product, form a Hilbert space $H$.
It is a standard result that a set of functions $\{ \psi_i \}$ 
is complete in $H$ 
when the linear span of the functions $\psi_i$ is dense
in $H$, and vice-versa (see, {\it e.g.}, \cite{cpt}).
This observation implies that the modes 
$\tilde{\psi}_i$
are complete in $L_2 ({\cal M})$.

We therefore have two complete and orthonormal sets of functions
$\psi_i$ and $\tilde{\psi}_i$ in $L_2 ({\cal M})$,
and an arbitrary field perturbation
$\delta \psi \in L_2 ({\cal M})$ can be 
expressed in terms of the modes $\tilde{\psi}_i$, {\it i.e.},
\be\label{owl2}
\ \delta \psi := \psi - \psi_0 = \sum_i \tilde{a}_i \tilde{\psi}_i.
\ee

It is simple to show that the 
transformation which expresses one set of 
basis functions
in terms of the other
must be orthogonal.
Let us now express the measure $\mb{d} [\psi]$,
given by expression (\ref{measure}), in terms of
the new set of modes $\tilde{\psi}_i$. We obtain,
\be\label{measure2}
\  \mb{d} [\psi] = \prod_{i}
\  \mu \int \mb{d}  \tilde{a}_{i} ,
\ee
where we used that the Jacobian of the transformation
relating the coefficients $a_i$ and $\tilde{a}_i$ equals  
one when the
transformation is orthogonal.

One could expect that the path-integral, evaluated 
with the measures (\ref{measure}) and (\ref{measure2}),
gives rise to the same result, 
since all we have done is to express 
one complete basis of modes in terms of the other.
This observation is not correct.
Note that when  
the path-integral (\ref{tau})
is performed with the measure (\ref{measure2}), 
then the source
term $J^+[ \psi]$ vanishes trivially, since
the argument $\psi$ is a linear combination 
of the modes $ \tilde{\psi}_i $, and
therefore $\psi \in \{ \delta \psi \}_0$. 
On the contrary, when the path-integral
is performed with the measure (\ref{measure}), then 
$\psi$ is a linear combination of the 
modes $ \psi_i $, and $J^+ [ \psi ]$ will generally be nonzero,  
which follows from the observation that 
$J^+[ \psi_i ]$ diverges for all $\psi_i$,
as we showed in the previous section.

Let us try to make precise in which sense the 
expansion
of the field in terms of 
two complete sets of modes 
(\ref{owl}) and (\ref{owl2}) differs.
Since both expansions converge to the same limit $\delta \psi$,  
it follows that the {\em difference} between the two
expansions can only be a configuration with zero $L_2 ({\cal M})$-norm.
When performing the path-integral (\ref{tau}),
using the measures (\ref{measure}) and 
(\ref{measure2}) respectively, we are 
integrating over paths 
in $L_2 ({\cal M})$ which may differ
by a zero-norm configuration.
These zero-norm configurations are precisely the degrees
of freedom which give rise to the nontrivial 
source term $J^+ [\psi]$.
In order to show this, let us recall that $J^+ [ \psi_i ]$ diverges for all 
$\psi_i$.
Since $J^+ [ \psi ]$ is linear in $\psi$,  
and $J^+ [ \tilde{\psi} ] = 0$ when $\tilde{\psi}$ is in the
linear span of the modes $\tilde{\psi}_i$, it
follows that 
\be\label{zero}
\ J^+ [ \psi_i ] = J^+ [ \psi_i - P \psi_i   ], 
\ee
where $P \psi_i$ denotes the projection of $\psi_i$ onto
the basis of modes $\tilde{\psi}_i$, {\it i.e.}, 
\be
\ P \psi_i := \sum_j  \langle \tilde{\psi}_j \psi_i \rangle \tilde{\psi}_j . 
\ee
But the modes $\tilde{ \psi}_i$ where found to be complete
in $L_2 ({\cal M})$, so that  
$(1 - P) \psi_i$ 
must have
zero $L_2 ({\cal M})$-norm.
The argument of $J^+$ on the right-hand side of equation
(\ref{zero}) has therefore zero $L_2 ({\cal M})$-norm,
and therefore this must be the degree of freedom which 
causes the divergence of the source term. 
Since the action functional depends on zero-norm degrees
of freedom through the term $J^+ [\psi]$, the expression
for the correlation function (\ref{tau}) 
is under-determined.
Recall that the same ambiguity was present when we tried 
to determine the extremal-action configurations 
in section \ref{osup}.
Although we do not know of a way to resolve this ambiguity,
let us consider two different approaches which might work. 

First, one can fix the zero-norm degrees of freedom on
the basis of a physical or philosophical argument. 
In practice, this could mean that one sets the source
term $J^+$ equal to zero by restricting the phase-space
of the $\psi$-field to a dense subset of $L_2 ({\cal M})$ for which 
$J^+$ vanishes.
In order to make this approach better than just guessing, one
needs to establish whether specific restrictions on the 
phase-space of the $\psi$-field lead to different predictions,
which can be falsified.

As a different approach, one could change the measure on 
the space of the $\psi$-field in order to accommodate 
the zero-norm degrees of freedom. Again, the problem
is that there is no clear guideline for doing so, 
unless one can show that different choices of 
measure lead to different observable predictions. 

It is illustrative to consider a similar ambiguity 
which occurs in the definition of the path-integral,
when one is dealing with fluctuations at 
infinitesimal rather than infinite length 
scales.  
This ambiguity is related to the fact that 
typical paths which contribute to the path-integral
are non-differentiable. Since the class of smooth
paths ($C^{\infty}$) is dense in the class of 
continuous paths ($C^0$), the difference between a path in $C^{0}$ 
and the nearest path in $C^{\infty}$
must have zero $L_2 ({\cal M})$-norm.
As we have seen, the measure (\ref{measure}) does not
accommodate these degrees of freedom, and the formal
expression is ambiguous on the point of the 
differentiability of the paths over which we integrate.
The action functional is however sensitive to the
degree of differentiability of the paths, 
which is made clear by the fact that  
the action is generally finite for differentiable paths
and infinite for non-differentiable paths. 
One could try to resolve this ambiguity   
by simply considering paths in $C^{\infty}$,
so that the action functional is well
defined, but in this case one can show
that the field operators in expression (\ref{tau})
commute trivially, and one does not recover
quantum physics \cite{path}.

Finally, let us note that similar implications hold 
for other field theories which are described by an
action functional which is non-linear in the 
field variable. 
In particular, it is well known that the Einstein
field equations can be derived by varying an action
functional, which is given by 
\be\label{dcx}
\ S [g_{\mu\nu}] = \fr{1}{16\pi G} \int_{\cal M} R \sqrt{-g},
\ee
where $R$ denotes the Ricci scalar, and 
we have ommited a possible contribution from matter fields and
a cosmological constant.
Similar to the case where we considered a scalar field, a contribution of
a surface term to the variation of the action  
does occur. At first-order in the metric
perturbation, the contribution of this surface term  
is given by \cite{waldo}, 
\be\label{dcm}
\ \delta S [g_{\mu\nu}] = -2 \int_{\partial {\cal M}} \left( \delta K 
\ + n^a h^{bc} \delta g_{ab;c} \right) \mb{d} \Omega,
\ee 
where $\delta K$ denotes the variation of the trace
of the extrinsic curvature at the boundary 
$\partial {\cal M}$, while $h^{bc}$ and $n^a$ denote
the induced three-metric and the normal to the
boundary respectively, and $\mb{d} \Omega$
denotes the volume element on $\partial {\cal M}$. 
The first term on the right-hand side of equation (\ref{dcm}) can be 
canceled by adding a surface integral of two times
the extrinsic curvature $K$ to the action functional
(\ref{dcx}) (see also the discussion in section 3).
The second term on the right-hand side of equation (\ref{dcm})
vanishes when it is evaluated according to a classical variational 
approach where we set $\delta g_{ab}$ equal to zero at the bounday
$\partial {\cal M}$, 
but this term could
be of interest in cosmological situations when we do
not require that perturbations vanish outside
a finite volume.

\section{Conclusion}

We revisited the variational principle in a cosmological context.
Starting from the path-integral formulation of quantum physics,
we argued that there is a correspondence between classical physics and
extremal action fields. The phase-space in which extremality of the
action is considered, is not constrained in quantum physics,
and we showed that there can be a non-trivial contribution
arising from surface terms.
We made this problem explicit by considering a scalar field in a
perturbed open FLRW spacetime. In the case of an open FLRW spacetime 
with a Gaussian
spectrum of subcurvature perturbations, we found no non-trivial
correction to the classical equation of motion.
In the case where supercurvature perturbations are present,
extremality of the action could only be defined after adopting
additional restrictions on the phase-space of the scalar field,
but the corresponding equations of motion are ambiguous since
they depend on how one restricts 
the phase-space of the field. 
We showed that the restricted phase-spaces which yield
different physical results, differ by perturbations
with vanishing $L_2$-norm.
This ambiguity is present both at a classical level and a
quantum level.  
We briefly discussed a possible strategy to resolve the 
ambiguity which is due to perturbations with 
vanishing $L_2$-norm.

\end{document}